\documentstyle[11pt,psfig,cite]{article}

\title{A two-dimensional network simulator for two-phase flow in porous media}
\author{Eyvind Aker\footnote{Also at: IKU Petroleum Research, 
N-7034 Trondheim, Norway}, Knut J\o rgen M\aa l\o y \\
Department of Physics \\ University of Oslo, N-0316 Oslo, Norway \\ \\
Alex Hansen\addtocounter{footnote}{-1}\footnotemark \\
Department of Physics, 
Norwegian University of Science  \\ and Technology, N-7034 Trondheim, 
Norway \\ \\
G. George Batrouni \\
Institut Non-Lin\'eaire de Nice, \\ 
Universit\'e de Nice - Sophia Antipolis, 06560 Valbonne, France}
\date{\today}

\begin{document}
\maketitle

\begin{abstract}
We investigate a two-dimensional network simulator capable of modeling
different time dependencies in two-phase drainage displacements. In
particular, we focus on the temporal evolution of the pressure due to
capillary and viscous forces and the time dependence of the interface
between the two liquids.  The dynamics of the capillary effect are
taken into account and we report on high accuracy pressure
measurements. Moreover, the simulator includes important features in
drainage, like burst dynamics of the invading fluid and simultaneous
flow of two liquids in a section of a tube. The validity of the model
is checked by comparing simulation results with well known experimental 
properties in drainage displacement.
\end{abstract}

\section{Introduction}
Two-phase displacements in porous media have been much studied
over the last two decades. The main reason for this is the great
variety of structures observed when changing the physical parameters
of the fluids like viscosity contrast, wettability, interfacial
tension and displacement rate. Besides being a process of great
interest in modern physics, it has a large number of practical
applications in many fields of science like oil recovery and
hydrology.

The main purpose of this paper is to present a network simulator
modeling immiscible two-phase flow on a two-dimensional lattice of
cylindrical tubes. Primarily, the model is developed to measure the time
dependence of different physical properties and to study the dynamics
of the fluid movements. We focus on drainage displacements, i.e.  the
process where a non-wetting fluid displaces a wetting fluid in a
porous medium.  In particular, we report on the dynamics of the
temporal evolution of the pressure due to capillary and viscous forces
and the time dependence of the interface between the two liquids.

The different structures between the invading and the defending fluids
obtained in drainage divide into three major flow regimes: viscous
fingering~\cite{Chen-Wilk85,Maloy85}, stable displacement~\cite{Len88}
and capillary fingering~\cite{Len85}. There exist statistical models
like DLA~\cite{Witten81}, anti-DLA~\cite{Pater84} and invasion
percolation~\cite{Wilk83} that reproduce the basic domains in 
viscous fingering, stable displacement and capillary fingering respectively.
However, these models do not contain any physical
time for the front evolution and the they cannot describe the
crossover between the major flow regimes.

To overcome the limitations of the statistical models several network
simulators similar to the one presented in this paper, have been
developed over the last two decades
\cite{Chen-Wilk85,Kop-Lass85,Paya86-1,King87,Len88,Blunt-King90,Paya96}.
The different models are more or less simplified to avoid
computational complications, and difficulties most often arise when
the capillary effects are taken into account.
Lenormand et al.\ (1988) assigned each tube a threshold
pressure and allowed the injecting non-wetting fluid to enter the tube
only when the pressure exceeded the threshold value of that
tube. Moreover, the non-wetting fluid was restricted to flow in the
positive direction relative to the direction of the
displacement. Dias and Payatakes (1986) suggested a more realistic
approximation by introducing tubes with walls with a sinusoidal
profile. They let the capillary pressure of the meniscus change as the
menisci invaded the tubes.

Similar to the idea of  Dias and Payatakes (1986)
the model reported here takes
into account the capillary effects by letting the capillary pressure
of the meniscus depend on its position inside the
tube. Moreover, the menisci can both invade into or retreat from a tube.
This is an important property in slow drainage, where the invading
fluid is found to suddenly invade a larger region causing another part
to retire (bursts)~\cite{Hain30,Len83,Maloy92}.  Experiments 
performed by M\aa l\o y et al.\ (1992) have also
produced evidence that the bursts are characterized by large pressure
fluctuations (Haines jumps). To measure these pressure
fluctuations the flow field in our network simulator is solved for a
constant injection rate rather than a constant pressure.

In the present model an approximation is developed to model
a mixture of the wetting and non-wetting fluids when they flow
simultaneously in a cross section of a tube. Such mixing is an
important process in both drainage and imbibition~\cite{Len83,Dull92} and
should not be neglected. The piston-like motion characterizing slow
drainage is well described, but additional mechanisms observed in
imbibition~\cite{Len83,Cieplak88} is not taken into account.

The paper is organized as follows. Section 2 describes 
the porous medium model used in the network simulator. Section 3
gives the solution of the flow field with the constraint of a constant 
injection rate, and Section 4 describes the algorithm
for updating the menisci. In Section 5 we discuss how
to move the menisci into neighboring tubes and allow mixing of 
the two liquids. Finally, in Section 6 we give some simulation result
focusing on calibrations of the model. At the end some conclusions 
are discussed.

\section{The Porous Medium Model}
\subsection{Geometry of Model Porous Medium}
The porous medium is represented by a square lattice of tubes inclined
45 degrees. Thus, if all tubes are equal and a uniform pressure
across the lattice is applied, a liquid flows equally well in tubes
inclined to the left as tubes inclined to the right. The tubes are
connected together at nodes, where four tubes meet. There is no volume
assigned to the nodes: the tubes represent the volume of both pores
and throats. The liquids flow from the bottom to the top of the lattice
and periodic boundary conditions are applied horizontally.  The
pressure difference between the first (bottom of system) and the last
(top) rows defines the pressure across the lattice.  Gravity effects
are neglected, and as a consequence we consider a horizontal flow in a
two-dimensional network of tubes. See figure~\ref{fig:network} for
details.

The tubes are cylindrical with equal length $d$, and every tube is
assigned a radius $r$ which is chosen at random in the interval
\mbox{$[\lambda_1,\lambda_2]d$}.  The randomness of the radii represent
the disorder of an ordinary porous medium and $\lambda_1$ and
$\lambda_2$ define the width of the distribution of radii.  

Initially, the system is filled with a defending fluid with viscosity
$\mu_1$. The invading fluid with viscosity $\mu_2$ is injected from the
bottom row with a constant injection rate.  
Here we report on the study of drainage displacements, and so let the invading 
fluid be non-wetting and the defending fluid be wetting. 
We assume that the fluids are immiscible and that there is a well defined 
interface between the two phases. The curvature of this 
interface gives rise to a capillary pressure given by the interfacial 
tension $\gamma$. Moreover, we will treat the liquids as incompressible,
i.e. the volume flux flowing into the system must equal the volume flux 
flowing out of the system.

\subsection{Capillary Flow in a tube}
\label{sec:tubeflow}
Consider a tube containing a meniscus between nodes $i$ and $j$ 
in the network as shown in figure~\ref{fig:tube}.
The volume flux $q_{ij}$ through the tube from the $i$th to the $j$th 
node is found from the Washburn equation for capillary flow~\cite{Wash21}:
\begin{equation}
q_{ij}=-\frac{\pi r_{ij}^2 k_{ij}}{\mu_{eff}}\cdot \frac{1}{d}(\Delta p_{ij} -p_{c})\ .
\label{eq:tubeflow}
\end{equation}
Here $k_{ij}=r_{ij}^2/8$ is the permeability, $\mu_{eff}=\mu_2 x_{ij}
+ \mu_1(1-x_{ij})$ is the effective viscosity due to the two
fluids, $\Delta p_{ij}=p_j-p_i$ is the pressure difference between the
$i$th and $j$th node and $p_c=p_1-p_2$ defines the capillary pressure.
$x_{ij}$ is the position of the meniscus in the tube and $r_{ij}$ is
the radius of the tube.  The position of the meniscus is a continuous
function in the range $[0,1]$ and the flow direction is given by the
sign of $q_{ij}$: When $q_{ij}>0$ the fluids in the tube flow to the
right, otherwise they flow to the left.  As mentioned in the
introduction, it is important to allow flow in both directions in order to
study the dynamics when the menisci invade into or retreat from a tube.

For a tube without a meniscus present $p_c=0$, and
equation~(\ref{eq:tubeflow}) reduces to that describing Hagen-Poiseulle
flow.

The capillary pressure  $p_c$ due to the meniscus is given by the
Young-Laplace law. Let $\theta$ refer to the wetting angle 
between the non-wetting and wetting
phases as shown in figure~\ref{fig:tube}. Then
\begin{equation}
p_c=\frac{2\gamma}{r}\cos\theta\ ,
\label{eq:pc}
\end{equation}
where $r/\cos\theta$ is the principal radius of curvature of the
meniscus. $\gamma$ denotes the interfacial tension between the two
phases.

Equation~(\ref{eq:pc}) is derived under the assumption that the fluids
are in static equilibrium, i.e. $\Delta p_{ij}=p_c$. A more detailed
description of the dynamics of a moving interface is still an open
problem~\cite{Duss79}. However, at low flow rates when no mixing or
turbulence occur, we consider the above expression as a reasonable
approximation.

A real porous medium consists of a complex network of throats and
pores with a great variety of shapes. The curvature of a meniscus
traveling in this network, becomes a continuous function depending on
the meniscus' position. The variation in curvature results in local
changes in the capillary pressure. Equation~(\ref{eq:pc}) does not take
this effect into account, since the capillary pressure is assumed
constant for each tube.  Instead, we apply a dependency in $p_c$ 
as a function of the menisci's position inside each
tube. Thus, we define the capillary pressure $p_c$ as (without
index notations):
\begin{equation}
p_c=\frac{2\gamma }{r}\left[ 1-\cos (2\pi x)\right]\ ,
\label{eq:pcvary}
\end{equation}
where we have assumed perfect wetting ($\theta=0$).  In the above
formula $r$ denotes the radius of the actual tube and $x$ is the
position of the meniscus in that tube. The function is plotted in
figure~\ref{fig:p_c}. The definition sets the capillary pressure equal
to zero at the ends of the tube whereas $p_c$ becomes equal to the
threshold pressure $p_t$ when the meniscus is in the middle of the
tube, i.e. $p_t=4\gamma/r$. The threshold pressure is the minimum
capillary pressure required to let the non-wetting fluid invade the
tube.

\section{Solving the Flow Field}
\label{sec:FlowField}
The fluids are assumed incompressible leading to conservation of volume flux
at each node:
\begin{equation}
\sum_j q_{ij}=0\ .
\label{eq:Kirch}
\end{equation}
Here $q_{ij}$ denotes the flow trough a tube connecting
node $i$ and $j$. Equation~(\ref{eq:Kirch}) is simply Kirchhoff's equation
and in the summation $j$ runs over the nearest neighbor nodes to the 
$i$th node. The index $i$ runs over all nodes that do not belong to the 
top or bottom rows, that is, the internal nodes. 
This set of linear equations are to be solved with the
constraint that the pressures at the nodes belonging to the upper and
lower rows are kept fixed.

By inserting equation~(\ref{eq:tubeflow}) in equation~(\ref{eq:Kirch}) we get:
\begin{equation}
\label{eq:Kirch_huge}
\sum_j G_{ij}(p_j-p_i-p_c)=0\ ,
\end{equation}
where $G_{ij}$ defines the mobility of the tube, i.e. 
$G_{ij}\equiv \pi r_{ij}^2 k_{ij}/\mu_{eff}$. In order to write this set of
equations as a matrix equation, we move all the capillary pressures and
the fixed pressure referring to the nodes belonging to the upper and
lower rows to the right-hand side of equation~(\ref{eq:Kirch_huge}). The
final matrix equation may then be written as
\begin{equation}
\sum_j D_{ij}p_j=B_i\ ,
\end{equation}
where the indices $i$ and $j$ only run over internal nodes. $D_{ij}$
are elements a conductance matrix $D$~\cite{CGM} where the elements
depend on the connection between different tubes and their
respective mobility.
$p_j$ are the elements in the pressure vector, containing the
pressure at the internal nodes and $B_i$ contains the pressure at the
boundaries (upper and lower rows) and the capillary pressure if a
meniscus is present in the tube.

We are seeking the solution of the pressure at the internal nodes for
a given configuration of the menisci, i.e
\begin{equation}
\label{eq:invKirch}
p_j=\sum_i {(D^{-1})}_{ij}B_i\ .
\end{equation}
This equation is solved by using the Conjugate Gradient method\newline 
\cite{CGM}.

\subsection{Solving for a Constant Injection Rate}
The pressures solved by equation~(\ref{eq:invKirch}) correspond to
keeping the applied pressure difference across the network constant.
We want to study the dynamics of the pressure fluctuations at a
constant displacement rate. Thus, we need to find the pressures $p_j$
under constant injection rate of the invading fluid.

For two-phase displacement in a porous medium the injection rate
$Q$ is given by
\begin{equation}
\label{eq:Darcy2}
Q=A\Delta P+B\ .
\end{equation}
Here $\Delta P$ is the pressure across the lattice and $A$ and $B$ are
parameters depending on the geometry of the medium and the 
current configuration of the liquids. The first part of 
equation~(\ref{eq:Darcy2}) is simply Darcy's law for one phase flow 
through a porous medium. The last part $B$ results from the capillary 
pressure between the two phases. As long as  
the menisci are kept at fixed positions $B$ becomes a constant.  

The pressure across the lattice for a given injection rate is
\begin{equation}
\Delta P = \frac{Q-B}{A}\ .
\label{eq:press}
\end{equation}
The parameters $A$ and $B$ are calculated by solving the pressure
field~(\ref{eq:invKirch}) for two different pressure $\Delta P^{'}$ and
$\Delta P^{''}$ applied across the lattice.  The obtained pressure
vectors give the corresponding injection rates, $Q^{'}$ and
$Q^{''}$. We are now in the position to calculate $A$ and $B$ 
by solving the set
\begin{eqnarray}
Q^{'}  & = & A\Delta P^{'}+B\ ,\\
Q^{''} & = & A\Delta P^{''}+B\ .
\end{eqnarray}

The next step is to relate the internal pressures at the nodes to the
constant injection rate. From equation~(\ref{eq:tubeflow}) we can
write the flow rate $q_{ij}$ from node $i$ to node $j$ like
\begin{equation}
q_{ij}=a_{ij}\Delta p_{ij}+b_{ij}\ ,
\label{eq:Darcy3}
\end{equation}
where $a_{ij}$ and $b_{ij}$ are parameters depending on the
permeability of the tube, the effective viscosity and the capillary
pressure due to a meniscus. From equation~(\ref{eq:press}) the
pressure $\Delta P$ for a given $Q$ is found. Thus, we could proceed
by solving Kirchhoff's equation with the constraint of applying this
pressure at the inlet. That would give the correct flow rates in each tube
for the desired $Q$. However, the following method will save a
third solution of equation~(\ref{eq:invKirch}). 

All the equations involved in the calculations are linear, i.e. they
have the functional form $f(x)=ax+b$. As a consequence the pressure
$\Delta p_{ij}$ between node $i$ and $j$ becomes a linear function of
the pressure $\Delta P$ across the lattice:
\begin{equation}
\Delta p_{ij}=\Gamma_{ij}\Delta P+\Pi_{ij}\ ,
\label{eq:pscale}
\end{equation}
Here are $\Gamma_{ij}$ and $\Pi_{ij}$ parameters depending on the
configuration of the fluids. By inserting this into
equation~(\ref{eq:Darcy3}) and redefine $a_{ij}\mbox{'s}$ and
$b_{ij}\mbox{'s}$ we obtain
\begin{equation}
\label{eq:qscale}
q_{ij}=\tilde{a}_{ij}\Delta P+\tilde{b}_{ij}\ .
\end{equation}
The parameters $\tilde{a}_{ij}$ and $\tilde{b}_{ij}$ are found from
the flow rates $q'_{ij}$ and $q''_{ij}$ corresponding to the two
pressures $\Delta P^{'}$ and $\Delta P^{''}$ respectively. Note that
the parameters $A$ and $B$ in~(\ref{eq:Darcy2}) and 
$\tilde{a}_{ij}$ and $\tilde{b}_{ij}$ in~(\ref{eq:qscale}) all
depend on the current position of the menisci and we therefore need to
solve them for every new fluid configuration. 

The solution due to a constant injection rate can now be summarized into two
steps: (a) After we have found $A$ and $B$ we use
equation~(\ref{eq:press}) to get $\Delta P$ for the desired $Q$. (b)
This $\Delta P$ is then used in equation~(\ref{eq:qscale}) to get the
local flow $q_{ij}$.

The validity of equation~(\ref{eq:pscale}) is easily checked by
solving Kirchhoff's equation for the calculated pressure $\Delta P$ and
compare the solution with the one given from equation~(\ref{eq:qscale}).
Numerical results show excellent agreement between these two
solutions.

\section{Updating the Menisci}
\label{sec:update}
A time step $\Delta t$ is chosen such that every meniscus is allowed
to travel at most a maximum step length $\Delta x_{max}$ during that time step.
This leads to the formula
\begin{equation}
\label{eq:timestep}
\Delta t=\min_{ij} \left[ \frac{\Delta x_{max}}{v_{ij}}\right]\ ,
\end{equation}
where $v_{ij}=q_{ij}/\pi r_{ij}^2$ denotes the flow velocity in a tube
containing a meniscus between the $i$th and the $j$th node. The time
step becomes dependent on the local velocity.  This method is
sometimes called event driven updating. 

During the time step it is checked whether or not a meniscus crosses
the middle or the end of a tube. If this happens, the time step is
redefined such that only one meniscus reaches the end or the middle of
the tube. A meniscus reaching the end of a tube is moved into the
neighbor tubes (see section~\ref{sec:move-neighbor}).

In the middle of the tube the capillary pressure becomes equal to the
threshold pressure. At this point, the menisci in slow drainage are
found to become unstable and suddenly invade the tube like a
burst~\cite{Hain30,Len83,Maloy92}.  Often a cascade of bursts are
released in rapid succession around the original instability, before a
new stable position is reached. Moreover, the bursts are characterized
by large pressure fluctuations~\cite{Maloy92}. It is important to
detect the order in which the instabilities occur to ensure that the
right path of least resistance is chosen.  This is done by adjusting
the time step when a meniscus approaches the middle of a tube such
that the exact occurrence of the burst is detected.

The new positions of the menisci are calculated by using a second
order Runge-Kutta scheme~\cite{NumRec}. Numerical analysis from the
present model shows that this scheme produces more stable solutions
than the less accurate and more unstable Euler scheme. However, with
an event driven time step as defined in equation~(\ref{eq:timestep}),
the implementation of the second order Runge-Kutta scheme is not
straight forward. Before discussing our modifications, we
first remind the reader of the Runge-Kutta scheme.

Let $x_n$ denote the position of a meniscus at time $t_n$ and
let the next position at time $t_{n+1}=t_n+\Delta t$ be $x_{n+1}$
where $\Delta t$ is the time step found from~(\ref{eq:timestep}).  The
second order Runge-Kutta scheme then becomes~\cite{NumRec}
\arraycolsep .07cm
\begin{eqnarray}
\label{eq:RK}
x_{n+1} & = & x_n +k_2 +O(\Delta t^3)\ , \\
\rule[.5cm]{0cm}{0cm}
k_1     & = & \Delta t\cdot v(t_n,x_n)\ , \\
\rule[.5cm]{0cm}{0cm}
k_2     & = & \Delta t\cdot v(t_n+\mbox{\Large $\frac{1}{2}$}\Delta t,x_n+\mbox{\Large $\frac{1}{2}$}k_1)\ ,
\end{eqnarray}
where $v(t_n,x_n)$ denotes the local flow rate in each tube and
$v(t_n+\frac{1}{2}\Delta t,x_n+\frac{1}{2}k_1)\equiv v_{mid}(t_n,x_n)$
defines the midpoint velocity.

Note that the velocities defining the time step in
equation~(\ref{eq:timestep}) correspond to the derivative of the curve
$x_n$ at the starting point of each time
interval. I.e. $v(t_n,x_n)=v_{ij}$, where the subscript $ij$ is
omitted on the left hand side of the equality. When the menisci are
updated by using the velocities $v(t_n,x_n)$,
equation~(\ref{eq:timestep}) leads to the condition
\begin{equation}
\label{eq:dxmx}
\left| x_{n+1}-x_n \right| \leq \Delta x_{max}\ ,
\end{equation}
for all the menisci. In the second order Runge-Kutta scheme the next
position $x_{n+1}$ is found by using the midpoint velocity
$v_{mid}(t_n,x_n)$, which in general differs from $v(t_n,x_n)$. As a
consequence, condition~(\ref{eq:dxmx}) may no longer be valid and the
displacement length or the stability of the numerical solution, are no
longer controlled. Assume that the position of the menisci is a smooth
function of time, the effect vanishes  since
$v_{mid}(t_n,x_n)\simeq v(t_n,x_n)$.  Moreover, there is no problem as
long as $v_{mid}(t_n,x_n)\leq v(t_n,x_n)$. The problem arises when
$v_{mid}(t_n,x_n)\gg v(t_n,x_n)$ and the final displacement becomes
much larger than $\Delta x_{max}$. To avoid this scenario the time
step is redefined by inserting the midpoint velocities in
equation~(\ref{eq:timestep}).  A new midpoint velocity corresponding
to the redefined time step is calculated and the positions are updated
according to this time step by using the new midpoint
velocity.  The procedure is repeated until the displacements become
close enough to the maximum step length $\Delta x_{max}$. Typically,
the final displacement must not be larger than about $10\%$ increase of 
$\Delta x_{max}$. Otherwise, the numerical solution may diverge.

\section{Motion of the Menisci at the Nodes}
\label{sec:move-neighbor}
A tube partially filled with one of the liquids is allowed to have
either one or two menisci leading to four different arrangements as
shown in figure~\ref{fig:nodeconf}. The menisci can be situated at any
position inside the tube and the corresponding effective viscosity and
capillary pressure are calculated.  The effective viscosity becomes
the sum of the fraction of the wetting and non-wetting fluid inside
the tube multiplied with their respective viscosities.  The absolute
value of the capillary pressure is given by
equation~(\ref{eq:pcvary}), while its sign depends on whether the meniscus
is pointing upwards like in figure~\ref{fig:tubeconf}\,(a) or downwards
like in figure~\ref{fig:tubeconf}\,(b). The menisci are updated
according to the determined time step $\Delta t$ from the previous
section, and their respective flow rates $v_{ij}$. The total time
lapse is recorded before the flow field is solved for the new fluid
configuration. Note that due to the incompressibility of the liquids
the two menisci within the same tube in figure~\ref{fig:tubeconf}~(c)
and (d) always move with equal velocities. Thus, the volumes of the
wetting in (c) or the non-wetting in (d) are conserved.

When a meniscus reaches the end of a tube it is moved into the
neighbor tubes according to some defined rules. These 
rules take care of the different fluid arrangements that can appear
around the node.  Basically, the non-wetting fluid can either invade
into or retreat from the neighbor tubes as shown in
figure~\ref{fig:nodeconf}\,(a) and (b) respectively.
In~\ref{fig:nodeconf}\,(a) the non-wetting fluid approaches the node
from below (drainage). When the meniscus has reached the end of the
tube (position 1), it is removed and three new menisci are created at
position $\delta$ in the neighbor tubes (position 2). The distance
$\delta$ is about 1--5\% of the tube length $d$ and it defines the
node region where the capillary pressure is zero. The node region
avoids that the created menisci at position 2 immediately disappear
and move back to the initial position 1 in tubes where the flow
direction is opposite to the direction of the invading fluid. The node
region has also an important role that allows mixing of
the liquids (see below).

Figure~\ref{fig:nodeconf}\,(b) shows the opposite case when the
non-wetting fluid retreats into a single tube (imbibition).  As
figure~\ref{fig:nodeconf} shows the properties of imbibition should
not be neglected as long as the menisci can travel in both
directions. However, in drainage which is what we are focusing on,
arrangement (b) will appear rarely compared to (a).

When the menisci are moved a distance $\delta$ into the neighbor
tubes the total time lapse is adjusted due to the injection rate
of the invading fluid. The adjustment is calibrated such that
the amount of the invading fluid in the lattice always equals the injected
volume. The injected volume is the product of the time lapse and the 
injection rate.

Difficulties arise when we want to create a new meniscus in a
tube that already contains two menisci. To allow such movement the
two original menisci and the new meniscus are merged into one
by the following method.  Consider a scenario like
figure~\ref{fig:nodeconf}\,(a) but assume now that the leftmost tube
already contains a bubble of non-wetting fluid as shown in
figure~\ref{fig:reduce}\,(a). To merge the three menisci in the leftmost
tube into one, we move the wetting fluid between position 2 and
3 to the left side of the non-wetting bubble and remove the menisci
at position 2 and 3 before the meniscus at position 4 is moved to the right 
a distance equal to the original length between position 2 and 3 
(figure~\ref{fig:reduce}\,(b)). The same principles apply when
the non-wetting fluid retreat into a tube that already contains a 
bubble of wetting fluid.

\subsection{Mixing of the Fluids}
\label{sec:mixing}
The moving rules described in figure~\ref{fig:nodeconf}
and~\ref{fig:reduce} solve the problem of modeling a ``mixture'' of
the two liquids. Consider the situations shown in figure~\ref{fig:mix}
where both the non-wetting and wetting fluids flow toward the node
from the bottom and right tube respectively. Physically, it is
expected that the fluid flowing out of the node is a mixture of both
liquids. It is observed~\cite{Len83,Dull92} that such simultaneous
flow of both liquids in a tube takes place in so-called funicular
continuous flow or a discontinuous dispersed flow. In the former case
the wetting phase flows along the cylinder wall surrounding the non-wetting
phase, which occupies the central portion of the tube. A discontinuous 
flow is characterized by a dispersed non-wetting phase flowing as
isolated droplets in a continuous wetting phase.  

A model describing funicular and dispersed flow would be too
complicated. Instead, we assume that the simultaneous flow can be
represented by a finite number of small bubbles of each liquid, placed
next to each other inside the tube. By sorting the bubbles of same
type of fluid they can be replaced by one or two menisci. The
procedure is illustrated with an example in figure~\ref{fig:mix}: When
the meniscus in the bottom tube reaches the end of the tube it moves a
distance $\delta $ into the neighbor tubes (arrangements a and b). Due
to the opposite flow direction in the right tube the created meniscus
in this tube flows back to the node (arrangement c) and moves into the
neighbor tubes (arrangement d). Now, the new meniscus in the bottom
tube again approaches the node from below (arrangement e) and creates
a configuration with three menisci in the top and left tubes
(arrangement f). To avoid that the number of menisci inside a single
tube increases unchecked the three menisci at the top and the left
tube are reduced to one by placing the wetting fluid on the top of the
non-wetting one (arrangement g).

The reorganization when three menisci 
are reduced to one (figure~\ref{fig:reduce}
and~\ref{fig:mix}) results in unphysical jumps in the capillary
pressure. Due to the small size of the bubbles, the jumps usually appear as
perturbations in the total pressure. A small number of such jumps
will not affect the numerical solution very much.  But, the moment
they become more dominant the numerical solution may become unstable
and diverge. We discuss this below.

\section{Calibration of the Network Model}
We present three different calibration simulations, one in each of the
regimes of interest: viscous fingering, stable displacement and
capillary fingering.  The simulations are executed on a Cray T90
vector machine.  The Conjugate Gradient method solving Kirchhoff's
equations, is easy to vectorize and achieves high performance
efficiency on vector machines. However, the amount of time required
increases dramatically with the size of the lattice.  For a lattice
consisting of $N$ tubes, the Conjugate Gradient method needs
approximately $N^2$ iterations each time step to solve Kirchhoff's
equations. Doubling the size of the lattice quadruples $N$, hence, the
computation time increases by a factor of $16$. In addition the
number of the time steps before the invading fluid reaches the outlet
strongly affects the CPU-time.

In two-phase fluid displacement there are mainly three types of forces:
viscous forces in the invading fluid, viscous forces in the defending
fluid and capillary forces due to the interface between them. This leads
to two dimensionless numbers that can characterize the flow in porous media: 
the capillary number $C_a$ and the viscosity ratio $M$.

The capillary number is a quantity describing the competition be\-tween 
capillary and viscous forces. It is defined as
\begin{equation}
C_a=\frac{Q\mu }{\Sigma\gamma }\ ,
\label{eq:Ca}
\end{equation}
where $Q$ ($\mbox{cm}^2/\mbox{s}$) denotes the injection rate, $\mu $
(Poise) is the maximum viscosity of the two fluids , $\Sigma$
($\mbox{cm}^2$) is the cross section of the inlet and
$\gamma $ (dyn/cm) is the interfacial tension between the two
phases. $\Sigma $ is the product of the length of the inlet and 
the mean thickness of the lattice due to the average radius of the tubes.

$M$ defines the ratio of the viscosities of the two fluids and is
given by the invading viscosity $\mu_2$ divided with the defending
viscosity $\mu_1$:
$$
M=\frac{\mu_2}{\mu_1}\ .
$$

The three simulations are performed with parameters as close as
possible to the corresponding experiments done
by M\aa l\o y et al.\ (1985)
and Frette et al.\ (1997). In light of that,
the length $d$ of all tubes in the lattices are set equal to
$1\,\mbox{mm}$.  Furthermore, the radii $r$ of the tubes are chosen
randomly in the interval $0.05d\leq r\leq d$.  The interfacial tension
is set to $\gamma=30\,\mbox{dyn/cm}$ and the viscosities of the
defending and the invading fluids varies between $0.01\,\mbox{P}$
($\simeq$ water) and $10\,\mbox{P}$ ($\simeq $ glycerol).

\subsection{Viscous Fingering}
\label{sec:VF}
Figure~\ref{fig:movie.vf} shows the result of a simulation in the
regime of viscous fingering performed on a lattice of $60\times 80$
nodes.  The corresponding pressure across the lattice as a function of
time is shown in figure~\ref{fig:press.vf}. The simulation is stopped
at breakthrough of the invading non-wetting fluid. The displacements are
done with a high injection rate, $Q=1.5\,\mbox{ml/min}$ and the
invading fluid is less viscous than the defending wetting
fluid, $\mu_2=0.010\,\mbox{P}$ and $\mu_1=10\,\mbox{P}$. The capillary
number and the viscosity ratio becomes $C_a=4.6\cdot 10^{-3}$ and
$M=1.0\cdot 10^{-3}$ respectively.

In viscous fingering the principal force is due to the viscous forces
in the defending fluid and the capillary forces at the 
menisci are less dominant. The pattern formation
(figure~\ref{fig:movie.vf}) shows that the invading fluid creates
typical fingers into the defending fluid. 

The pressure across the lattice (figure~\ref{fig:press.vf}) decreases
as the less viscous fluid invades the system. Roughly, the pressure
appears to decrease linearly as a function of time. However, the slope
is non-trivial and results from the fractal development of the
fingers. In addition to the fractal growth, the rate
of change in the pressure depends on the viscosity contrast $M$ between
the two phases. The rapid decrease at the end of the pressure function
corresponds to the breakthrough of the invading fluid at the outlet.

Figure~\ref{fig:press.vf} shows that there are small fluctuations in the
average decreasing pressure function. The fluctuations
correspond to the chang\-es in the capillary pressure as a meniscus
invades into or retreats from a tube. The fluctuations are small compared
to the total pressure and thus the capillary pressure cannot affect the
result very much.

In addition to containing important physics, the pressure function in
figure~\ref{fig:press.vf} is used to establish the stability and
convergence properties of the numerical solution. For all simulations
it is verified that this function converges towards a unique solution
when $\Delta x_{max}\leq 0.1d$ and the menisci are updated according
to the second order Runge-Kutta method as described in
section~\ref{sec:update}. That means Kirchhoff's equation must be
solved approximately 10--20 times to let a meniscus pass through a
single tube. This is probably what we can expect when we want to
measure the fluctuations in the pressure due to local capillary
changes inside the tubes.  With larger step lengths the variations in
the capillary pressure are lost and the solution is no longer
suitable for our measurements.

\subsection{Stable Displacement}
Figure~\ref{fig:movie.sd} shows the result of a simulation performed
in the regime of stable displacement on a lattice of $60\times 60$
nodes.  Figure~\ref{fig:press.sd} shows the corresponding pressure
across the lattice as a function of time. The simulation is stopped
when the invading fluid has about half filled the system.  As in the
case of viscous fingering $Q=1.5\mbox{ ml/min}$ and $C_a=4.6\cdot
10^{-3}$. The viscosities are $\mu_1=0.10\mbox{ P}$ and $\mu_2=10\mbox{
P}$ giving $M=1.0\cdot 10^2$. Thus, the invading non-wetting fluid is
more viscous than the defending wetting fluid.

The fluid movements are dominated by the viscous forces in the
invading liquid and like viscous fingering the capillary effects are
negligible. The invading fluid generates a compact pattern with an
almost flat front between the non-wetting and wetting fluid. The simulation
is stopped after the width of the front has stabilized, that means after
the width of the front stops growing.

The average pressure across the lattice (figure~\ref{fig:press.sd})
increases according to the amount of the high viscosity invading fluid
injected into the system. Due to the low viscosity defending fluid the
pressure corresponds to the pressure across the invading phase and a
linear increase in the pressure is observed after the front has
stabilized. (In figure~\ref{fig:press.sd}: $t>50\,\mbox{s}$). Like in
viscous fingering, the average slope of the pressure function also
depends on the viscosity ratio $M$ and the injection rate.

The viscous forces dominate the pressure evolution, but fluctuations
due to capillary effects are observed. The perturbations have about
the same size as for viscous fingering, which is not surprising since
the size distribution of the radii of the tubes is the same for the
two cases. The threshold pressures setting the strength of the
capillary fluctuations, is inversely proportional to the radius of the
tubes.

In fast displacement the viscous forces are strong enough to deform
and even move small regions of trapped defending fluid which are left
behind the front. It is also observed that the invading fluid grows
along the whole front, causing the corresponding menisci to
reach the end of the tube approximately at the same time. All together
the result is a lot of bubbles of non-wetting and wetting fluid which
travel through the network. Physically, the bubbles are expected but
at the time they become too dominant in the displacement they cause some
computational complication. The bubbles increase the number of tubes
that should contain three menisci. Thus, a large number of reorganizations 
inside the tubes are applied to fulfill the constraint of at most 
two menisci in each tube. Typically, this is observed in
figure~\ref{fig:press.sd} where the fluctuations in last part of the
pressure function seem to increase in amplitude. The increasing
amplitude is a result of the unphysical pressure jumps when three
menisci are reorganized into one. For that reason the
simulation is stopped before breakthrough of the invading fluid.

Another shortcoming in fast displacement is that the effective time
step may approach zero. When the bubbles develop the number of menisci
increases causing many of them to arrive at the end of the tubes
approximately at the same time. The result is that the difference of
the arrival times for the menisci goes to zero and a lot of iterations
are required to move the whole front further on. Numerical experiments
have shown that the problem first arises when the displacement is
rather fast, that means $C_a>5\cdot 10^{-3}$.

\subsection{The Regime of Capillary Fingering}
Finally a calibration simulation is run in the regime of capillary
fingering. The resulting pattern is shown in figure~\ref{fig:movie.ip},
and in figure~\ref{fig:press.ip} the corresponding pressure across the
lattice as a function of time is plotted. The simulation is performed
on a lattice of $40\times 60$ nodes and the run is stopped when the
invading fluid reaches the outlet. The two fluids have equal viscosities,
$\mu_1=\mu_2=0.50\mbox{ P}$, corresponding to similar experiments performed 
by Frette et al.\ (1997). The invading non-wetting fluid
displaces the defending wetting fluid with a rather low injection
rate, $Q=0.20\mbox{ ml/min}$.  The capillary number and the viscosity
ratio become $C_a=4.6\cdot 10^{-5}$ and $M=1.0$ respectively.

In capillary fingering the displacement is so slow that the viscous
forces are negligible, with the consequence that the main force is the
capillary one between the two fluids.  Only the strength of the
threshold pressure in the tubes decides if the invading fluid 
invades that tube or not. Since the radii of the tubes (which determine
the threshold pressures) are randomly chosen from a given
interval, the non-wetting fluid flows a random path of least
resistance. 

Figure~\ref{fig:movie.ip} shows a typical rough front
between the invading and the defending fluids with trapped clusters of
defending fluid left behind the front. As opposed to stable
displacement the clusters appear at all sizes between the tube length
and the maximum width of the front.

The pressure across the lattice (figure~\ref{fig:press.ip}) exhibits
sudden jumps according to the capillary variation when the non-wetting
fluid invades (or retreats) a tube. The fluctuations identify the
bursts where the invading fluid proceeds abruptly. An enlargement of a
small part of the pressure function at at time around $550\,\mbox{s}$
is given in figure~\ref{fig:burst}. The figure shows clearly this kind
of dynamics. The pressure across the lattice slowly increases in
stable periods before the threshold pressure in the tube which is
going to be invaded is reached.  At the threshold pressure the
meniscus becomes unstable and the invasion of fluid takes place in a
burst accompanied by sudden negative jumps in the pressure. The
pressure curve in figure~\ref{fig:burst} is in good qualitative
agreement with experimental results~\cite{Maloy92}.

The main problem in capillary fingering is the computation time. The
above simulation used about $36$ CPU hours at a Cray T90, on a smaller
lattice than the ones for viscous fingering and stable displacements.
The two latter cases required both about $7$ CPU hours. The reason for
this dramatic increase in computation time is explained by looking at
the physical properties of capillary fingering. The invasion occurs in
bursts and each burst is localized to some very few tubes. That means
that even with step lengths equal to those of stable displacement the
amount of invading fluid injected into the system each time step is
much less. Now, compared to viscous fingering the total saturation of
the invading fluid is relatively high and as a consequence, an
enormous number of time steps are required to reach break through.  In
the above simulation approximately $300,000$ time steps are applied,
while in the case of viscous fingering only about $2,500$ steps were
necessary to reach breakthrough.

\section{Conclusions}
We have presented and discussed a two-dimensional network model
simulating drainage displacement. Moreover, we have performed
numerical calibration simulations whose results are found to be in
good qualitative agreement with the properties observed in the three major
flow regimes: viscous fingering, stable displacement and capillary
fingering.

An important feature of the model is the capability to study temporal
evolutions of different physical properties taking place in the
displacement process. This continues the work of understanding
two-phase flow in porous media. In particular, we have presented calculations
of the pressure evolutions where the dynamics of the capillary effects
were taken into account. With the proposed model it is also possible to
measure the viscous and capillary contribution to the total
pressure. This is a great advance since corresponding measurements in
experimental setups are too difficult. A quantification of
the competition between viscous and capillary forces will help to
characterize the different flow regimes observed in drainage displacement.

An approximation has been developed to simulate a simultaneous flow of
the two liquids inside a single tube. The presented moving rules take
care of this kind of mixing and actually, they are essential to make
the model work. So far the moving rules are based on the mechanisms
observed in drainage, but in principle it should be possible by a
certain modifications to model imbibition as well.

The lattice sizes are limited by the computation time and the
model requires usage of high performance vector machines.  To afford
simulations on lattice sizes comparable to those in experimental
setups~\cite{Maloy85,Inge97}, more sophisticated and efficient
algorithms have to be developed. Especially, vector machines with
parallel capabilities should be considered.

\section*{Acknowledgments}
We thank S.\ Basak, I.O.\ Frette and J.\ Schmittbuhl for valuable
comments. The computations were done at HLRZ, For\-schungs\-zent\-rum
J\"{u}\-lich GmbH and E.A.\ is grateful for the hospitality shown
there. The work has received support from NFR, The Research Council
of Norway.

\begin{figure}
\begin{center}
%\mbox{\psfig{figure=Figures/network.eps,height=3.5cm}}
\caption{A square lattice of tubes connected together at nodes. The size
of the lattice is $10\times 10$ nodes. The black region indicates the 
invading non-wetting fluid coming from below and the light gray indicates 
the defending wetting fluid flowing out of the top.}
\label{fig:network}
\end{center}
\end{figure}
\begin{figure}
\begin{center}
%\mbox{\psfig{figure=figs/tube.eps,height=2.5cm}}
\caption{Flow in a tube containing a meniscus.}
\label{fig:tube}
\end{center}
\end{figure}
\begin{figure}
\begin{center}
%\mbox{\psfig{figure=figs/pcap.eps,height=4cm}}
\caption{The capillary pressure $p_c$ as a function of the meniscus'
position $x$ in that tube. In the middle of the tube at 
$x=d/2$ the capillary pressure becomes equal to the threshold pressure $p_t$.}
\label{fig:p_c}
\end{center}
\end{figure}
\begin{figure}
\begin{center}
%\mbox{\psfig{figure=figs/fluidconf.eps,height=2.5cm}}
\caption{Four different fluid arrangements inside one tube. The shaded
and the white regions indicate the non-wetting and wetting fluid 
respectively.}
\label{fig:tubeconf}
\end{center}
\end{figure}
\begin{figure}
\begin{center}
%\mbox{\psfig{figure=figs/nodeconf.eps,height=3cm}}
\caption{The motion of the menisci at the nodes. (a): The non-wetting
fluid (shaded) reaches the end of the tube (position 1) and is moved a
distance $\delta$ into the neighbor tubes (position 2). (b): The
wetting fluid (white) reaches the end of the tubes (position 1) and
the non-wetting fluid (shaded) retreat to position 2.  For both (a) and
(b) a proper time is recorded due to the small movement $\delta $.}
\label{fig:nodeconf}
\end{center}
\end{figure}
\begin{figure}
\begin{center}
%\mbox{\psfig{figure=figs/reduce.eps,height=3cm}}
\caption{Reduction of three menisci into one. (a): The
non-wetting fluid has reached the end of the tube (position 1) and is
going to be moved into the neighbor tubes (position 2). The three
menisci in the left tube are reorganized causing the
situation shown in (b) to appear. In the figures the arrows denote the
length of the wetting fluid between position 2 and 3 which is equal to
the distance the meniscus at position 4 is moved to the right.}
\label{fig:reduce}
\end{center}
\end{figure}
\begin{figure}
\begin{center}
%\mbox{\psfig{figure=figs/mix.eps,height=8.0cm}}
\caption{A ``mixture'' of non-wetting (shaded) and wetting (white)
which flow into the neighbor tubes. The different arrangements
(a)--(g) are a result of applying the rules which are described
earlier in this section. For all figures the fluids flow towards the
node from the bottom and right tube while the fluids in the top
and the left tube flow away from it (denoted by the arrows in (a)).}
\label{fig:mix}
\end{center}
\end{figure}
\begin{figure}
\begin{center}
%\mbox{\psfig{figure=Figures/movie.vf.eps,height=6cm}}
\caption{The pattern obtained of a  simulation in the regime of viscous
fingering on a lattice of $60\times 80$ nodes.  $C_a=4.6\cdot 10^{-3}$
and $M=1.0\cdot 10^{-3}$. The invading non-wetting fluid (black)
displaces the defending wetting fluid (gray) from below. The
simulation took about $7\mbox{\tiny $\frac{1}{2}$}$ hours on a Cray T90 vector
machine.}
\label{fig:movie.vf}
\end{center}
\end{figure}
\begin{figure}
\begin{center}
%\mbox{\psfig{figure=Graphs/press.vf.eps,height=6cm}}
\caption{The calculated pressure across the lattice as a function of
time for viscous fingering. The pressure decreases with time due to the
viscous forces in the defending fluid and the flow velocities of the
moving finger tips.  The perturbations correspond to the capillary
forces due to the moving menisci. The time is the total time
lapse required to let the invading fluid reach the outlet.}
\label{fig:press.vf}
\end{center}
\end{figure}
\begin{figure}
\begin{center}
%\mbox{\psfig{figure=Figures/movie.sd.eps,height=5.2cm}}
\caption{The pattern obtained of a simulation performed on a lattice of
$60\times 60$ nodes in the regime of stable displacement.
$C_a=4.6\cdot 10^{-3}$ and $M=1.0\cdot 10^{2}$. The invading
non-wetting fluid (black) displacement the defending wetting fluid
(gray) from below. The simulation took about $7$ hours on a Cray T90
vector machine.}
\label{fig:movie.sd}
\end{center}
\end{figure}
\begin{figure}
\begin{center}
%\mbox{\psfig{figure=Graphs/press.sd.eps,height=6cm}}
\caption{The calculated pressure across the lattice as a function of
time for stable displacement. For $t>50\,\mbox{s}$ the pressure
increases linearly with time due to the viscous forces in the invading
fluid and the constant injection rate.  The perturbations correspond
to the capillary forces along the front.}
\label{fig:press.sd}
\end{center}
\end{figure}
\begin{figure}
\begin{center}
%\mbox{\psfig{figure=Figures/movie.ip.eps,height=6cm}}
\caption{The pattern obtained of a simulation performed on a lattice
of $40\times 60$ nodes in the regime of capillary fingering.
$C_a=4.6\cdot 10^{-5}$ and $M=1.0$. The invading non-wetting fluid
(black) displacement the defending wetting fluid (gray) from
below. The simulation took about $36$ hours on a Cray T90 vector
machine.}
\label{fig:movie.ip}
\end{center}
\end{figure}
\begin{figure}
\begin{center}
%\mbox{\psfig{figure=figs/press.ip.eps,height=6cm}}
\caption{The calculated pressure across the lattice as a function of
time for capillary fingering. The fluctuations due to the capillary
forces correspond to the burst dynamics of the invading fluid. See
also figure~\protect\ref{fig:burst}.}
\label{fig:press.ip}
\end{center}
\end{figure}
\begin{figure}
\begin{center}
%\mbox{\psfig{figure=figs/burst.eps,height=6cm}}
\caption{A magnification of the pressure across the lattice as a function of 
time in the interval $540\mbox{--}560\,\mbox{s}$. The fluid invasion takes 
place in bursts characterized by the the negative pressure jumps.}
\label{fig:burst}
\end{center}
\end{figure}

\end{document}